\documentclass[aps,prl,twocolumn,showpacs,preprintnumbers,amsmath,amssymb]{revtex4}
\usepackage{graphicx}% Include figure files
\usepackage{dcolumn}% Align table columns on decimal point
\usepackage{bm}% bold math
\newcommand{\nc}{\newcommand}
\nc{\be}{\begin{equation}}
\nc{\ee}{\end{equation}}
\nc{\bea}{\begin{eqnarray}}
\nc{\eea}{\end{eqnarray}}
\nc{\bean}{\begin{eqnarray*}}
\nc{\eean}{\end{eqnarray*}}
\nc{\mb}{\mbox}
\nc{\rnc}{\renewcommand}
\nc{\vk}{\mb{\bf k}}
\nc{\vp}{\mb{\bf p}}
\nc{\vn}{\mb{\bf n}}
\nc{\vq}{\mb{\bf q}}
\nc{\rr}{\mb{\bf r}}
\nc{\vz}{\hat {\mb{\bf z}}}
\nc{\vj}{\mb{\boldmath$j$}}
\nc{\vg}{\mb{\boldmath$g$}}
\nc{\x}{\mb{\boldmath$x$}}
\nc{\A}{\mb{\boldmath$A$}}
\nc{\va}{\mb{\boldmath$a$}}
\nc{\vs}{\mb{\boldmath$\sigma$}}
\nc{\vpi}{\mb{\boldmath$\pi$}}
\nc{\nab}{\nabla}
\nc{\X}{\sf x}

\begin{document}

%\preprint{APS/123-QED}

\title{Quantum Transport of Massless Dirac Fermions}
\author{Kentaro Nomura}
\affiliation{Department of Physics, University of Texas at Austin,
Austin TX 78712-1081, USA}
\author{A. H. MacDonald}
\affiliation{Department of Physics, University of Texas at Austin,
Austin TX 78712-1081, USA}

\date{\today}

\begin{abstract}
 Motivated by recent graphene transport experiments,
 we have undertaken a numerical study of the conductivity of 
 disordered two-dimensional massless Dirac fermions.  
 Our results reveal distinct differences between the cases of 
 short-range and Coulomb randomly distributed scatterers.
 We speculate that this behavior is related to the 
 Boltzmann transport theory prediction of dirty-limit behavior 
 for Coulomb scatterers. 
 \end{abstract}
 
 \pacs{72.10.-d,73.21.-b,73.50.Fq}
% PACS, the Physics and Astronomy
                             % Classification Scheme.
%\keywords{Suggested keywords}%Use showkeys class option if keyword
                              %display desired
\maketitle

\noindent
{\em Introduction}---
Graphene can be described at low-energies by a four component massless Dirac-fermion (MDF)
model\cite{slonczewski}  that has long attracted theoretical attention 
because of appealing properties including chiral anomalies\cite{jakiew,semenoff,fradkin1,haldane},
randomness induced quantum criticality\cite{fradkin2,ludwig,chamon96},
unusual electron-electron interaction effects\cite{kim97,khveshchenko01},
relevance to high $T_c$ superconductors\cite{lee,kim97}, and various unusual transport properties\cite{haldane,fradkin2,ludwig,lee,morita,ando,neto,graphene_qhet,sheng,hatsugai93,chamon96,kim97,
minimalc,khveshchenko01,suzuura,localization}. 
The recent experimental realization of
single-layer graphene sheets\cite{geim} has made it possible to confirm a number
of theoretical predictions, including unusual quantum Hall effects\cite{graphene_qhe1,graphene_qhe2}, 
and has also revealed some surprises.  The main findings can be summarized as follows:
i) graphene's conductivity $\sigma$ never falls below a minimum
value ($\sigma^{\rm min}$) corresponding (approximately) to a conductance
quantum ($e^2/h$) per channel, in spite of predictions\cite{fradkin2,ludwig,lee,ando,neto}
based on the self-consistent Born approximation (SCBA) that $\sigma^{\rm min} = (1/\pi) \, e^2/h$ for a MDF channel,
and predictions that localization occurs\cite{suzuura,localization,minimalc} when intervalley scattering is significant;
ii) in gate-doped graphene $\sigma$ increases linearly with 
carrier density $n$ away from the charge neutrality (Dirac) point, implying a constant mobility 
$\mu=\sigma/ne$\cite{graphene_qhe1,graphene_qhe2} and 
not the constant conductivity usually predicted for the Boltzmann transport regime\cite{ando}. 
Although these surprises have inspired a number of theoretical studies\cite{neto,localization,minimalc},
the source of the discrepancies between experiment and theory has not yet been conclusively identified. 

We have recently pointed out that the linear dependence of conductivity on carrier density in graphene 
can be explained in the framework of Boltzmann transport theory by assuming Coulomb scatterers\cite{nomura} 
rather than the short-range scatterers assumed, mainly as a practical simplification, in most theoretical work.
The golden-rule scattering rate for Coulomb scatterers in a MDF model 
diverges in the zero energy (Dirac point) limit, whereas it vanishes for short-range scatterers.
This property suggests the possibility of a qualitative difference between these 
two disorder models.  Since Boltzmann theory is not applicable in the
vicinity of the Dirac point, however, a fully quantum mechanical approach is required to address the minimal conductivity.
In this Letter we report on a finite-size Kubo formula analysis used to study the quantum transport of MDFs.
Over the range of system sizes that we can describe, we find that $\sigma^{\rm min}$ is a few times larger 
for Coulomb scatterers than for short-range scatterers and $\sim e^2/h$. 
 
A MDF model describes graphene transport only when intervalley scattering is unimportant.
We argue that the MDF model likely does apply at accessible experimental temperatures 
to graphene systems near the Dirac point, since the intervalley scattering length obtained within the Born approximation diverges
in that limit.  The temperature below which intervalley scattering is important 
should increase away from the Dirac point\cite{graphene_localization}.

\noindent
{\em Massless Dirac Fermion Model}---
Graphene's honeycomb lattice has two atoms per unit cell on 
sites labeled $A$ and $ B$. 
The low-energy band structure consists of Dirac cones located at
the two inequivalent Brillouin zone corners $K$ and $K'$:
\bea
H_{\rm K}= 
 \hbar \, v \, \vs\cdot\vk  = \ \ 
v \, \hbar \left(
\begin{array}{rr}
 0\ \ \ \ \   &  k_x-ik_y   \\
 k_x+ik_y  &  0\ \ \ \ \  
\end{array}
\right)\; ,
\label{band}
\eea
and $H_{\rm K'}=\hbar v\vs^t\cdot\vk$,
where $v$ is the graphene Fermi velocity and the Pauli matrices $\vs$ act on the sublattice
degrees of freedom.   For each wavevector $\vk$,
Eq.(\ref{band}) has two eigenstates $|\vk,\pm\rangle=(|\vk,A\rangle\pm
e^{i\phi}|\vk,B\rangle)/\sqrt{2}$, where $\phi\equiv\tan^{-1}(k_y/k_x)$,
and eigenenergies $E_{\vk,\pm}=\pm \hbar v |\vk|$.  
When intervalley scattering is neglected the $K$ and $K'$ valleys
contribute independently to the conductivity.

\begin{figure}[!t]
\begin{center}
\includegraphics[width=0.4\textwidth]{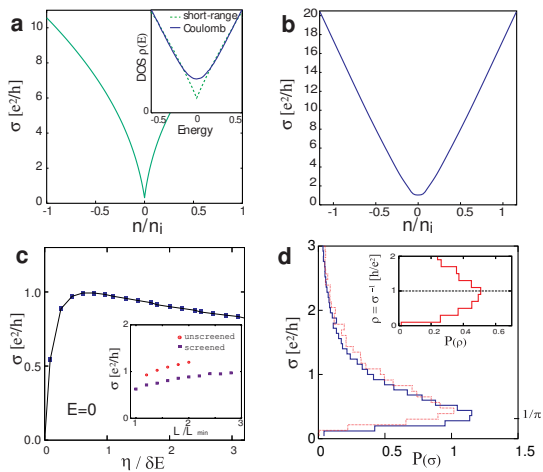}
\caption{
(Color online) Dirac fermion conductivities for (a) short-range scatterers and 
(b) screened Coulomb scatterers. The inset of (a) compares the densities of states for short-range and
 Coulomb cases. 
  (c) shows Kubo formula conductivities at the Dirac point  as a function of $\eta$ for Coulomb scatterers for three
 different system sizes.
The inset of (c) shows the size dependence of the conductivity for unscreened Coulomb (open circles) and 
screened Coulomb (boxes) cases.(d) Distribution function of the conductivity and resistivity (inset) at the Dirac point for
Coulomb scatterers. In the main panel, the solid (blue) line is for $L=1.4L_{\rm min}$ 
and the dotted (red) line is for $L=2L_{\rm min}$, where $L_{\rm min}$
is the minimum system size considered.
}
\label{diagrams}
\end{center}
\end{figure}

\noindent
{\em Boltzmann theory for doped graphene}---We start by briefly reviewing 
Boltzmann transport theory applied to graphene since this consideration motivates
Colomb scatterer models.  The Boltzmann conductivity 
$
\sigma_0^{s} = (e^2/h)\;  (2 E_F \tau_0/\hbar) =(e^2/h)\; 2k_F\ell
$
is proportional to the transport relaxation time $\tau_0$.
For short-range scatterers the Born
approximation gives $\hbar/\tau_0=2\pi{\overline
{V^2}}\rho_F=(n_iu^2/\hbar v^2) \; |E_F|\; $\cite{ando}, where
$V(\rr)=u\sum_I^{N_i}\delta(\rr-{\bf R}_I)$ is the
disorder potential, $n_i$ is the density of
scatterers, and $\rho_F$ the density of states 
at the Fermi level.  When the range of the impurity potential is much longer than the lattice
spacing of graphene, intervalley scattering is weak\cite{ando,suzuura,localization}.

Note that $\sigma_{0}^s$ is independent of the carrier density $n$. 
Experiment, on the other hand, finds that the mobility $\mu =  \sigma / ne$ 
in graphene is nearly constant except at very low-densities.
One plausible explanation for this behavior is that Dirac fermion scattering
is dominated by Coulomb scattering from ionized impurities near the
graphene plane, $V(\rr)=\sum_I^{N_i}e^2/\epsilon |\rr-{\bf R}_I|$.
Using Fermi's golden rule and approximating the screened Coulomb 
interaction by\cite{nomura} $U_{\rm sc}({\bf q})=(2\pi e^2)/\epsilon (q+4\alpha_g k_F) \simeq (\hbar v \pi)/(2 k_F)$, 
the Boltzmann conductivity for Coulomb interactions is 
$\sigma_{0}^c \simeq (4e^2/h) \; (n/n_i)\; 32/\pi$, proportional to density in agreement 
with experiment.  Here $\alpha_g = e^2/\epsilon \hbar v$ 
is the effective fine structure constant used to characterize the ratio of Coulomb interaction 
and band energy scales in graphene.
 ($\alpha_g\simeq 3$ in vacuum and $\simeq 1$ when the graphene sheet is placed on a ${\rm Si}{\rm O}_2$ 
dielectric substrate.) Note that the Boltzmann conductivity for Coulomb scatterers vanishes as 
$n\rightarrow 0$, contradicting experiment\cite{graphene_qhe1}.

\noindent
{\em MDF model finite-size Kubo Formula --}
Our numerical results, obtained by evaluating the finite-size Kubo formula 
\bea
 \sigma =-\frac{i\hbar e^2}{L^2}\sum_{n,n'}\frac{f(E_{n})-f(E_{n'})}{E_{n}-E_{n'}}
\frac{\langle n|v_x|n'\rangle\langle n'|v_x|n\rangle}{E_{n}-E_{n'}+ i\eta},
\label{kubo}
\eea
are summarized in Fig.1.  Here ${\bf v}=v\vs$ is the Dirac fermion velocity operator, $f(E)$ is the
Fermi-Dirac function, and $|n\rangle$ denotes an eigenstate 
of the Dirac equation,
\bea
\label{dirac}
 \left[-i\hbar v \vs\cdot\nabla+V(\rr)\frac{}{}\right]\psi=E\psi
\eea 
which we solve using a large momentum-space cutoff $\Lambda$.
The disorder potential momentum-space matrix elements are
\bea
\langle \vk\sigma|V|\vk'\sigma'\rangle=\frac{1}{L^2}\sum_{I=1}^{N_i}U(\vk-\vk')\delta_{\sigma\sigma'}\ e^{i(\vk-\vk')\cdot {\bf R}_I},
\eea
where $U({\bf q})$ is constant for delta-impurities and given by $U_{\rm sc}({\bf q})$ for the screened Coulomb scattering case.
The scattering center locations ${\bf R}_I$ and potential signs were chosen at random. 
The results shown in Fig.1 are typically averaged over $10^{4}$ disorder realizations.
We estimate the bulk conductivity by evaluating Eq.(\ref{kubo}) at a large number of
$\eta$-values.  The Kubo conductivity $\sigma$ vanishes for both small and large $\eta$ but there
is an intermediate region where the dependence of $\sigma$ on $\eta$ is relatively weak.
We use the maximum of $\sigma$ {\em vs.} 
$\eta$ to estimate the conductivity at a given system size $L$.
For metals, including doped graphene, physical arguments suggest that
$ \eta \sim \hbar/ T_L$ where $T_L$ is the escape time from the system 
studied numerically. 
It follows\cite{thouless} that 
$\eta \sim \langle\Delta E\rangle \sim g_T \, \delta E$ where
the Thouless energy $\langle\Delta E\rangle$ is the geometric mean of the
eigenvalue difference between periodic and antiperiodic boundary
conditions, $\delta E=1/L^2\rho_F$ is the level spacing at the Fermi level, and $g_T^{}\equiv\langle\Delta E\rangle/\delta E$ is the Thouless number.

The reliability of the finite-size Kubo formula method
described above is solidly established for diffusive metallic systems\cite{thouless}. 
%For example typical results obtained when the methods employed here for the MDF model are applied to a conventional (non-relativistic) two-dimensional electron system are shown in Fig.1(c)
For conventional non-relativistic two-dimensional electron systems with parabolic dispersion, numerical 
conductivities calculated in this way agree closely with the theoretical expectation
of a Drude conductivity $\sigma_0= (e^2/h)k_F \ell$ with a finite-size weak-localization correction\cite{lee-review}.  
Subtleties, related to the $\eta$ dependence of the numerical estimate, 
do however arise when this method is applied to insulators. 
The Dirac point of the MDF model is an intermediate case and 
related uncertainties apply to our numerical estimate of $\sigma^{\rm min}$. 
The property that the Dirac point density of states is finite in the presence of disorder,
as illustrated in the inset of Fig.1(a), may help validate the Kubo approach.
We find that the dependence of $\sigma$ on $\eta$ at the Dirac point, illustrated 
in Fig.1(c) is similar to that in a metal\cite{thouless}, indicating delocalized states.
$\sigma$ is a maximum when $\eta$ corresponds to $\sim \delta E$ at each system size 
as expected for $\sigma \sim e^2/h$.
We have also evaluated the Thouless conductivity as a consistency check.
A large number of numerical studies have demonstrated that the Thouless
conductivity estimate, although perhaps not as accurate in the diffusive metal
limit ($ \sigma >> e^2/h$), is more universally applicable.  It may be used for 
both delocalized and strongly localized states\cite{thouless-qhe},
and should therefore be reliable at the Dirac point.  We find consistency between the two $\sigma$ estimates.
%which supports our conclusions.  

Fig.1 compares Kubo conductivity estimates for (a) 
the short-range scatterer and (b) the screened Coulomb scatterer cases. 
For the short-range disorder potential case, the density dependence of
the conductivity is non-linear, approaching the constant Boltzmann conductivity
for $ |E_F| \gg \hbar /\tau$.
The estimated value of $\sigma({E=0})$ is close to the SCBA value 
$(1/\pi)e^2/h$
predicted in earlier theoretical studies\cite{fradkin2,ludwig,lee,ando,neto}.
For the screened Coulomb scatterer case, on the other hand, the conductivity $\sigma$
increases linearly with increasing density $|n|$ as Boltzmann theory predicts.
At the Dirac point, however, the conductivity remains finite with the
minimum value $\simeq e^2/h$, which is few times larger than the SCBA value for the short-range model.  
To illustrate the dependence on carrier density, we have smoothed the curves in Fig.1(a) and (b) by 
averaging over Fermi energy interval containing typically 1-30 levels, and over boundary conditions, in addition to over approximately $10^4$ disorder realizations.  

If we neglect intervalley scattering
and account for graphene's spin and valley degeneracies, 
MDF properties can be compared with graphene experimental results.  
Both the linear dependence of the conductivity on density and the shift of $\sigma^{\rm min}$ suggest that 
long-range scattering similar to that produced by ionized impurities is present in experimental samples.
We note that ionized impurities in the substrate or at samples edges that are separated from 
the conduction channel by a distance that is large compared to the graphene lattice constant but small
compared to the Fermi wave length in the regime studied experimentally ($|n|\lesssim 7\times 10^{12} {\rm [cm^{-2}]}$\cite{graphene_qhe1,graphene_qhe2}) will act like Coulomb scatterers in the MDF model but will not produce strong intervalley scattering.

The difference between Coulomb and short-range
$\sigma^{\rm min}$ values can be rationalized by the following argument.
In the short-range case, the golden-rule relaxation time for delta-function scatterers
diverges ($\tau_0\propto 1/|E_F|\rightarrow\infty$) at the Dirac point and 
$k_F \ell$ remains finite, which suggests that standard transport theory considerations
may apply at least qualitatively once the non-vanishing density of states in disordered systems is accounted for.
In contrast, Boltzmann transport theory suggests that the 
Dirac point of the Coulomb scatterer model is in the strong disorder limit because $\tau_c\propto |E_F|\rightarrow 0$.
We examined this idea by varying the carrier density $n$ and
the ionized impurity density $n_i$ independently, finding that the conductivity for Coulomb scattering model appears to be  
a function of $n/n_i$ only.  Letting $n\rightarrow 0$ at finite $n_i$ and letting $n_i\rightarrow \infty$ at finite 
$n$ are equivalent because $\sigma\rightarrow \sigma^{\rm min}\sim e^2/h$. 
This finding is consistent with the idea that at
the Dirac point MDF Coulomb scatterers are always in the strongly disordered limit.
As long as these states remain delocalized, however, the conductivity cannot vanish;
the non-zero conductivity at the Dirac point is purely due to
the quantum mechanical nature of delocalized states.
Although this argument is not quantitative, it
makes it clear that there is an essential difference between the 
Coulomb and short-range cases.  
A related difference is also seen in the inset in Fig.1(a) which compares densities of states
near the Dirac point ($E=0$) for short-range scatterers (green
dashed line) and Coulomb scatterers (blue solid line).
The prominent dip at $E=0$ in the short-range case is replaced by a smooth minimum at 
a larger value in the Coulomb case.
One interpretation of the increase in Dirac point density of states 
is that the carrier density fluctuates spatially in the smooth
Coulomb potential.  Non-zero local carrier densities at the Dirac point could explain both the 
increase in density of states and the increase in conductivity.
For short-range scatterers, our simulations have focused on the Boltzmann dominated
$k_F\ell >> 1$ regime.  We note that systems with $k_F \ell \sim 1$ or smaller
behave more like Coulomb scatterer systems at low densities, and have larger 
minimum conductivities and densities of states. 

Although our theory does not account for electron-electron interactions, the 
issue of screening effect near the Dirac point requires comment.
We use the $T=0$ Thomas-Fermi approximation $U(q)=2\pi e^2/\epsilon (q+8\pi e^2\rho_F)$ to 
give an indication of the importance of screening, although this approximation must fail for 
$E_F \to 0$ since it applies strictly only for disorder potential range 
much longer than the Fermi wavelength and $k_BT \ll E_F$. 
The density of states at the Dirac point $\rho_F$ vanishes in the clean limit 
whereas it can be evaluated self consistently in the disordered case, leading to the small but finite value seen in the inset in Fig.1(a).
The inset in Fig.1 (c) compares the conductivity for unscreened and screened case at several different system sizes.
We find that the conductivities evaluated with screened interactions are suppressed somewhat compared to 
the unscreened case.  It appears likely that the precise value of $\sigma^{\rm min}$ may depend indirectly
on electron-electron interactions.

We emphasize that we are able to evaluate the Kubo formula only over a relatively small range of finite  
system sizes.  If we use the mobility to convert to physical length units
the minimum system size we study is $\sim 0.1 {\rm \mu m}$, not much 
smaller than the size of the crystallites studied experimentally.\cite{graphene_qhe1}
Over the range of sizes we are 
able to study the conductivity does increase
very slowly with system size, even at $E=0$, as shown in the inset of Fig.1(c).
This weak size dependence might be hinting that the infinite conductivity predicted for 
two-dimensional systems with symplectic symmetry\cite{2Dsymplecscaling} by scaling 
arguments will emerge in the MDF model at extremely large system sizes even at $E=0$.
It is not obvious to us that the scaling theory applies at the Dirac point for 
either Coulomb or short-range scatterers since there 
is no length scale on which $\sigma \gg e^2/h$.  Alternately, the weak size dependence
we find might be related to enhancement of carrier-density spatial fluctuations, which can be imagined 
in terms of puddles of electrons and holes, at larger system sizes in the long-range disorder potential case.
Since Dirac fermions are not strongly localized, these puddles are not isolated but effectively percolate\cite{klein} 
and are constrained by boundary condition in finite size systems. 
The large length scale limit of the conductivity at zero temperature is more obvious for real graphene than 
for its (two-component) MDF model, since intervalley scattering will always become relevant 
and lead to localization.\cite{suzuura,localization,minimalc}

Finally we comment on statistical properties of the conductivity and the resistivity at the Dirac point. Figure.1 (d) shows the distribution function of the conductivity (main panel) and the resistivity (inset). The distribution function of $\sigma$ has a sharp peak near $\sigma=(1/\pi)e^2/h$, and a large tail on the the large $\sigma$ regime. On the other hand, the resistivity distribution function has a broad peak near $h/e^2$, as seen in experiment.\cite{graphene_qhe1}

\noindent
{\em Discussion --} 
We find that a MDF model with Coulomb scatterers is able to account
for two key findings of experiments on graphene sheets, namely that the 
conductivity is proportional to carrier density away from the Dirac point and 
that the minimal conductivity per channel is finite and  
larger than the SCBA value obtained theoretically for a MDF model with  
short-range scatterers.  The impurity density $n_i$ in the Coulomb model 
should be associated with the density of ionized impurities that are located
in the substrate within a Fermi wavelength of the graphene plane. 
In this interpretation, the mobilities measured in 
current samples\cite{graphene_qhe1,graphene_qhe2} correspond to 
$n_i\simeq 5\times 10^{11}$ [cm$^2$].
The property that the conductivity is  
proportional to carrier density in the Boltzmann regime suggests dominant 
smooth intravalley scattering that could be Coulomb in character.
%Although intervalley scattering does not appear to be important in the 
%Boltzmann regime, it is much more likely to be relevant for weak 
%localization corrections at low temperature.
If so intervalley scattering is likely to be irrelevant in 
graphene in the Boltzmann transport regime, although 
it will always be important at finite carrier densities in the weak localization
regime when the temperature is low enough that the phase coherence
length exceeds the intervalley scatterling length, $l_0\propto 1/|E_F|$.
When intervalley scattering is weak, the MDF model applies and momentum space Berry phases change
constructive interference of back scattering particles into destructive interference\cite{ando2,suzuura}, 
implying weak antilocalization.  When local scatterers dominate in graphene intervalley and intravally 
scattering rates\cite{ando,suzuura,localization} are comparable and the accumulated Berry phases are randomized ,
so that all states are localized.\cite{morita,sheng}  Qualitative studies of electron-phonon interaction\cite{phonon} 
and electron-electron interaction\cite{khveshchenko01} in graphene give important information on the coherence length. 
In a weak magnetic field, the magnetic length limits the coherence length.
One recent magnetoresistance study has shown that highly doped graphene exhibits a negative magneto resistance 
indicating weak-localization, while states remain delocalized in the vicinity of the Dirac point \cite{graphene_localization}.
%In the presence of mesoscopic ripples in samples, the quantum corrections have been argued to be strongly suppressed.  These issues come up with various %interesting theoretical questions.
These properties further support the view that graphene is described by a MDF model near
the Dirac point.  Our numerical results suggest that the MDF conductivity is finite and 
$\sim e^2/h$ at the Dirac point over the relevant range of the system size and the coherence length. 

{\em Acknowledgment --}
The authors acknowledge helpful interactions with A. Geim, T. Hughes, Z. Jiang, J. Jung, P. Kim, V. Falko 
and A. Castro-Neto. This work has been supported by the Welch Foundation and by the Department of Energy under grant
DE-FG03-02ER45958.

{\em Note added --} After submitting the present paper, we became aware of related work in which the importance of the long-range nature of disorder scatterers is discussed.\cite{add1}

\end{document}